\documentclass[]{article}

\usepackage{amsmath}
\usepackage{bm}
\usepackage[dvips,xdvi]{graphicx}

\newcommand{\qstar}{{q^{*}}}
\newcommand{\transpose}[1]{#1^{t}}

\begin{document}


\title{\bf Density Functional Theory for Block Copolymer Melts and Blends}
\author{ Takashi Uneyama${}^{1}$ and Masao Doi${}^{2}$ \\
\\
${}^{1}$ Department of Computational Science and Engineering, \\
Graduate School of Engineering, Nagoya University, \\
Chikusa, Nagoya, JAPAN \\
${}^{2}$ Department of Applied Physics, \\
Graduate School of Engineering, The University of Tokyo, \\
Hongo, Tokyo, JAPAN}
\date{}
\maketitle

%
%

\begin{abstract}
We derive an expression for the free energy of the blends 
of block copolymers expressed
as a functional of the density distribution of the monomer of each block. 
The expression is a generalization of the Flory-Huggins-de Gennes theory
for homo polymer blends, and also a generalization of the Ohta-Kawasaki theory
for the melts of diblock copolymers.  The expression can be used for any 
blends of homopolymers and block copolymers of any topological structure.
The expression gives a fast and stable computational method to calculate the
micro and macro phase separation of the blends of homopolymers and block copolymers.
\end{abstract}


%
%

\section{Introduction}

Block copolymers and their blends show various interesting micro structures 
in equilibrium due to the difference in the segmental interaction between
the polymer blocks \cite{Bates-Fredrickson-1999}.  The phenomena has been extensively
studied by the self consistent
field (SCF) theory
\cite{Helfand-Wasserman-1976,Matsen-Schick-1994,Matsen-2003,Fraaije-1993,Zvelindovsky-vanVlimmeren-Sevink-Maurits-Fraaije-1998,vanVlimmeren-Maurits-Zvelindovsky-SevinkFraaije-1999,Drolet-Fredrickson-1999,Fredrickson-Ganesan-Drolet-2002,Kawakatsu-book,Matsen-Bates-1996}
in which the free energy of the
system is calculated by evaluating the path integral (or by solving the
Edwards equation) for each polymers
in the mean field.  This method is quite useful as it can 
calculate the free energy of any blends of polymers
of arbitrary topological structure. 
On the other hand, the SCF calculation is computationally demanding as it
needs large memory and large CPU power.

Many works have been done to reduce the CPU time and the memory needed 
for the SCF calculation.  If the system has a periodic structure of known
symmetry, the equilibrium structure (or the local equilibrium structure
for the given symmetry) can be  calculated rather efficiently by obtaining 
the eigen functions for the Edwards equation \cite{Matsen-Schick-1994,Matsen-2003}, 
or by using the narrow interface
approximation and the unit cell approximation \cite{Helfand-Wasserman-1976,Matsen-2003}. On the other hand,if the
system does not have periodic structure (such as in the case of micellar
systems) or if the symmetry of the system is not known, other strategies
are needed to get equilibrium structures.  Fraaije and his coworkers
developed a theory for the dynamics and conducted several dynamic
simulations, such as the dynamic behavior of the block
copolymer melts under shear flow or the aqueous solution of the block copolymers
\cite{Fraaije-1993,Zvelindovsky-vanVlimmeren-Sevink-Maurits-Fraaije-1998,vanVlimmeren-Maurits-Zvelindovsky-SevinkFraaije-1999}.
Fredrickson et. al. proposed an efficient SCF algorithm for real
space simulation \cite{Drolet-Fredrickson-1999}, and 
introduced complex fields to improve the efficiency
of convergence \cite{Fredrickson-Ganesan-Drolet-2002}.

Though significant progress has been made in the SCF calculation, 
it is still very difficult to simulate the 3 dimensional large systems of reasonable size of polymers
as the required computational resources increase dramatically with the increase of the
chain length: for 3 dimensional simulation, the calculation is practically limited to the 
system where the effective degree of polymerization is not so large.

An alternative approach is the density functional (DF) theory first
introduced by Leibler \cite{Leibler-1980}.
In this approach, the free energy $F$ of the system
is expressed as a functional of the density distribution function
of each monomer species. Clearly 
the computational cost of such approach is much less than that
of the SCF calculation since the evaluation of the path integral which costs
the major part of the cpu time and memory in the SCF calculation is not
needed.  
For example, consider to simulate the block copolymer solutions: the
degrees of polymerization of solvents are unity and ones of block
copolymers are large. We cannot assume any periodicity for such a
systems, thus we need to do the real space simulation.
The computational cost of the SCF calculation increases dramatically as the
degrees of polymerization of block copolymers increase, while the cost of
the DF simulations is independent of the degree of polymerization.

In the first DF theory \cite{Leibler-1980} applied to AB type diblock copolymers,  
the density functional was obtained as a power series of 
$\delta \phi(\bm{r})=\phi(\bm{r}) - \bar{\phi}$, 
where $\phi(\bm{r})$ is the volume fraction of A segment at point $\bm{r}$,
and $\bar{\phi}$ is the spatial average of $\phi(\bm{r})$.
\begin{equation}
  \bar{\phi} = \frac{1}{V}  \int d\bm{r} \phi(\bm{r})
\end{equation}
where $V$ is the volume of the system. The
free energy is expressed as \cite{Kawakatsu-book}
\begin{equation}
 \label{intro_1}
\begin{split}
 & F \left[ \phi(\bm{r}) \right] = F \left[ \bar{\phi} \right] \\
 & \qquad + \frac{1}{2}  
     \int d\bm{r} d\bm{r}' \, 
       \Gamma(\bm{r} - \bm{r}')   
         \delta\phi(\bm{r}) \delta\phi(\bm{r}') \\
 & \qquad + \frac{1}{3!}
     \int d\bm{r} d\bm{r}' d\bm{r}'' \, 
       \Gamma^{(3)}(\bm{r} - \bm{r}',\bm{r} - \bm{r}'')   
         \delta\phi(\bm{r}) \delta\phi(\bm{r}') \delta\phi(\bm{r}'') \\
 & \qquad + \dotsb \\
\end{split}
\end{equation}
where $F \left[ \bar{\phi} \right]$ is the free energy for homogeneous state.
The vertex functions $\Gamma(\bm{r})$, and 
$\Gamma^{(3)}(\bm{r},\bm{r}')$ are calculated by solving the
Edwards equation or by calculating the density correlation functions. 

Leibler's method is mathematically sound, but it can be applied  only for the
the case of weak segregation, i.e., the
case that the deviation from the homogeneous state is small.  
In order to apply the DF theory to the strong segregation
case, Ohta and Kawasaki 
proposed an approximate expression for the free energy \cite{Ohta-Kawasaki-1986,Ohta-Kawasaki-1990}.
They noticed that in  
Leibler's theory,
the term which is essential for giving the
micro phase separation is the ``long range'' term. This term
appears in the limit of small wave vector $\bm{q}$ in
the Fourier transform of $\Gamma(\bm{r})$ 
\begin{equation}
 \label{intro_2}
 \Gamma(\bm{q}) 
    = \int  d\bm{r} e^{-i \bm{q} \cdot \bm{r}} \Gamma(\bm{r})
\end{equation}
For small wave vector $\bm{q}$, $\Gamma(\bm{q})$
diverges as follows:
\begin{equation}
 \label{intro_3}
 \Gamma(\bm{q})=  \frac{A}{\bm{q}^2}  \qquad \text{for} \quad \bm{q}^{2} \to 0
\end{equation}
where $A$ is a certain constant.  The divergence comes from the fact that
the monomer A of block copolymers are always connected to monomer B and therefore
cannot separate from monomer B further than the  size of the block
polymer.
Ohta and Kawasaki then proposed the following form of the
free energy functional for the diblock polymer:
\begin{equation}
 \label{intro_4}
 F \left[\phi(\bm{r})\right] = 
 F_{L}\left[\phi(\bm{r})\right] + F_{S}\left[\phi(\bm{r})\right]    
\end{equation}
where $F_{L}\left[\phi(\bm{r})\right]$ and  $F_{S}\left[\phi(\bm{r})\right]$ stands for the free energy due to the 
long range and short range interactions.  The long range part is expressed as
\begin{equation}
 \label{intro_5}
   F_{L}\left[\phi(\bm{r})\right] =
   \int d\bm{r} d\bm{r}' \, \frac{A}{2} \mathcal{G}(\bm{r} - \bm{r}') 
   \delta\phi(\bm{r}) \delta\phi(\bm{r}')
\end{equation}
where $\mathcal{G}(\bm{r})$ is the inverse Fourier transform of 
$1 / \bm{q}^{2}$.
\begin{equation}
 \label{intro_6}
    \mathcal{G}(\bm{r}) = \frac{1}{(2 \pi)^{d}}
    \int d\bm{q} e^{i \bm{q} \cdot \bm{r}} \frac{1}{\bm{q}^2}
\end{equation}
where $d$ is the dimension of the space. In the real space,
$\mathcal{G}(\bm{r} - \bm{r}')$ satisfies
\begin{equation}
 - \nabla^{2} \mathcal{G}(\bm{r} - \bm{r}') = \delta(\bm{r} - \bm{r}')
\end{equation}

As to the short range part, Ohta and Kawasaki assumed the Cahn-Hilliard free
energy:
\begin{equation}
 \label{intro_7}
   F_{S}\left[\phi(\bm{r})\right] =
     \int d \bm{r} \left[ \frac{B}{2} \left|\nabla \phi(\bm{r})\right|^2 + f\left( \phi(\bm{r}) \right) \right]
\end{equation}
where $B$ is a constant and $f\left(\phi(\bm{r})\right)$ is a certain function which has double
minima.

As the Ohta-Kawasaki theory involves a few intuitive arguments,
generalization of the theory to other systems was not straightforward.  
Indeed the
generalization has been done only for limited systems such as
the ABC triblock copolymers
\cite{Nakazawa-Ohta-1993,Ren-Wei-2003b},
the blends of AB diblock copolymer and  C homopolymer \cite{Ohta-Ito-1995}, 
and the blends of A homopolymer,  B homopolymer and
AB diblock copolymer \cite{Kawakatsu-1994}.

In this paper we propose a general expression for the free energy which can be
applied for any types of block copolymers and their blends. All parameters
in the free energy are expressed by the microscopic parameters appearing
in the SCF theory, i.e., the structure of block copolymers and the 
interaction parameters (the $\chi$ parameters).
This expression gives a fast computational
method to calculate the equilibrium structure formed by the blends of block
polymers. 

An approach similar to ours was already taken by Bohbot-Raviv and Wang
\cite{BohbotRaviv-Wang-2000}: they
proposed an approximate expression for the free energy functional 
which can be applied for arbitrary structure of block polymers. 
We will compare our theory with their theories later.

The present paper is constructed as follows. 
Firstly, we obtain the free energy functional 
for the situation that the deviation from the homogeneous state is small. 
This step is rather straightforward: the free energy is
obtained by using the standard procedure of the linearized mean
field approximation, often referred to as the random phase approximation (RPA) \cite{deGennes-book,Kawakatsu-book}.
Secondly we generalize this expression and  
seek an expression which can be applied for the situation that the
density deviation from the homogeneous state is large.
The second step is rather arbitrary, but we 
found an expression which reduces to the Flory-Huggins-de Gennes theory 
for homopolymer blends \cite{deGennes-1980} and to Ohta-Kawasaki theory for the melts of
diblock copolymers. Thirdly we test this expression of the free energy  
by comparing the results of this expression with that of the SCF theory, 
and discuss the validity of the expression.
%
%

\section{Free Energy Functional for Small Density Perturbation}
\subsection{The Second Order Vertex Function}
We consider a mixture of homo polymers
and block copolymers which consist of several types of monomer units 
A, B, C.... The chemical structure of block copolymers are characterized by
the species of monomer units (such as A,B,C...),  the number of monomers in 
each block, and the connectivity of these blocks. 
Polymers having the same chemical structure, i.e., polymers consisting of 
the same blocks (of same monomer units and of same length) connected
in the same way are regarded to belong to the same polymer type. 
The polymer type is distinguished by the suffices $p,q,r,\dots$ , and
each block in the polymer is distinguished by the suffices,
$i,j,k,\dots$ (see Figure~\ref{mean_square_distance}).

Let $N_{pi}$ be the number of monomers belonging to the 
the $i$-th block of polymer $p$.
The total number of monomers in the polymer $p$ is given by
\begin{equation}
  N_{p} = \sum_{i} N_{pi}
\end{equation}
The block-ratio of the $i$-th block  of polymer $p$ is defined by
\begin{equation}
  f_{pi} \equiv \frac{N_{pi}}{N_{p}}
\end{equation}

We assume that all monomer units A,B,.. have the same specific volume and 
same bond length (this assumption can be removed but here it is assumed 
for the  sake of simplicity). 
Let $\phi_{pi}(\bm{r})$ be the volume fraction of the monomer
belonging to the block $(p,i)$ at point $\bm{r}$.  Our objective
is to find out the free energy expressed as a 
functional of $\{ \phi_{pi}(\bm{r})\} $ .

We first consider the situation that the system is 
homogeneous at equilibrium, and calculate the change of the free energy
for small density variation.
In the homogeneous state, $\phi_{pi}(\bm{r})$ is equal to
\begin{equation}
 \bar{\phi}_{pi} = f_{pi} \bar{\phi}_{p}.
\end{equation}
Let $\delta \phi_{pi}(\bm{r}) = \phi_{pi}(\bm{r}) - f_{pi} \bar{\phi}_{p} $
be the deviation of the monomer distribution from the homogeneous state.
The free energy functional for the system can be expressed as a quadratic
form of $\delta \phi_{pi}(\bm{r})$
\cite{Leibler-1980,Ohta-Kawasaki-1986}.
\begin{equation}
 \label{freeenergy_gl_real}
 \begin{split}
  & F \left[\{\phi_{pi}(\bm{r})\}\right] = \, F \left[\{f_{pi} \bar{\phi}_{p}\}\right] \\
  & \qquad + \frac{1}{2} \sum_{pi,qj} \int d\bm{r} d\bm{r}' \, \Gamma_{pi,qj}(\bm{r} - \bm{r}') \delta\phi_{pi}(\bm{r}) \delta\phi_{qj}(\bm{r}') \\
  & \qquad + \dotsb
 \end{split}
\end{equation}
where $F \left[\{f_{pi} \bar{\phi}_{p}\}\right]$ is the free energy for
the homogeneous state, and 
$\Gamma_{pi,qj}(\bm{r} - \bm{r}')$ is the second order vertex
function. 
In the Fourier space, eq~\eqref{freeenergy_gl_real} can be written as follows 
\begin{equation}
 \label{freeenergy_gl_wavenumber}
 \begin{split}
  & F \left[\{\phi_{pi}(\bm{q})\}\right] = \, F \left[\{f_{pi} \bar{\phi}_{p}\}\right] \\
  & \qquad + \frac{1}{2} \sum_{pi,qj} \frac{1}{(2 \pi)^d} \int d\bm{q} \, \Gamma_{pi,qj}(\bm{q}) \delta\phi_{pi}(-\bm{q}) \delta\phi_{qj}(\bm{q}) \\
  & \qquad + \dotsb
 \end{split}
\end{equation}
where $\delta \phi_{pi}(\bm{q})$ and $\Gamma_{pi,qj}(\bm{q})$ are the 
Fourier transform of $\delta \phi_{pi}(\bm{r})$ and $\Gamma_{pi,qj}(\bm{r} )$:
\begin{align}
 \delta\phi_{pi}(\bm{q}) 
 = & \int d\bm{r} \, e^{-i \bm{q} \cdot \bm{r}} \delta\phi_{pi}(\bm{r}) \\
 \Gamma_{pi,qj}(\bm{q}) 
 = & \int d\bm{r} \, e^{-i \bm{q} \cdot \bm{r} } \Gamma_{pi,qj}(\bm{r})
\end{align}

The second order vertex function $\Gamma_{pi,qj}(\bm{q})$ determines the density
fluctuation in the homogeneous state and can be
related to the density correlation function $S_{pi,qj}(\bm{q})$ which
is defined by
\begin{equation}
 \label{S_def} 
 S_{pi,qj}(\bm{q})
   = \left \langle \delta\phi_{pi}(\bm{q}) \delta\phi_{qj}(-\bm{q}) \right \rangle
\end{equation}
where $\langle \dots \rangle$ stands for the ensemble average for the
equilibrium state.  The density correlation function $S_{pi,qj}(\bm{q})$ 
is related to $\Gamma_{pi,qj}(\bm{q})$ by the following equation:
\begin{equation}
 \label{S_G_ortho}
 \sum_{q,j} \Gamma_{pi,qj}(\bm{q}) S_{qj,rk}(\bm{q}) = \delta_{pr}\delta_{ik}
\end{equation}

For ideal chains for which there is no interaction between 
monomer units, the density correlation arises from the connectivity
of the polymer chain, and exists only for the 
monomers belonging to the same polymer chain. Hence
$S_{pi,qj}(\bm{q})$ for the ideal chain is written as: 
\begin{equation}
 \label{S_ideal}
 S^{(\text{ideal})}_{pi,qj}(\bm{q}) = \bar{\phi}_{p} \delta_{pq}h_{p,ij}(\bm{q}) 
\end{equation}
where $h_{p,ij}(\bm{q})$ is the correlation function for a single
chain.  Let $\bm{r}_{pi}^{(s)}$ be the position of the $s$-th monomer in the
$i$-th block of the polymer $p$, then $h_{p,ij}(\bm{q})$ is given by
\begin{equation}
 \label{h_def}
 h_{p,ij}(\bm{q}) = \frac{1}{N_p}\int_0^{N_{pi}} ds \, \int_0^{N_{pj}} ds' \,
            \left \langle 
               \exp \left[ 
                        i \bm{q} \cdot (\bm{r}_{pi}^{(s)}-\bm{r}_{pj}^{(s')}) 
                    \right] 
          \right \rangle
\end{equation}
For Gaussian chain, the average can be calculated easily to give
\begin{equation}
 \label{correlation_ideal}
 h_{p,ij}(\bm{q}) = 
  \begin{cases}
   \displaystyle 
     \frac{2 N_{p} f_{pi}^{2}}{\xi_{pi}^{2}} 
       \left( 
             e^{-\xi_{pi}} - 1 + \xi_{pi} 
       \right) & (i = j) \\
   \displaystyle 
       \frac{N_{p} f_{pi} f_{pj}}{\xi_{pi} \xi_{pj}} 
       \left( 
          e^{-\xi_{pi}} - 1 
       \right)
    \left( 
         e^{-\xi_{pj}} - 1 
    \right) e^{-l_{p,ij}^{2} \bm{q}^{2} / 6}
    & (i \neq j)
   \end{cases}
\end{equation}
where 
\begin{equation}
 \label{xi_def}
 \xi_{pi} = \frac{1}{6}N_{p} f_{pi} b^{2} \bm{q}^{2} 
\end{equation}
$b$ is the effective bond length and
\begin{equation}
 \label{m_pij_def}
 l_{p,ij}^{2} = M_{p,ij} b^{2}
\end{equation}
Here  $M_{p,ij}$ is the number of monomers
included in the shortest path connecting the 
$i$-th block and the $j$-th block of the
$p$-th polymer, i.e., the chemical distance between the $i$-$j$ blocks.

Thus for ideal chains,  the correlation function 
can be calculated by eqs~\eqref{S_ideal} and \eqref{correlation_ideal}, 
and the vertex function can be calculated by eq~\eqref{S_G_ortho}. 
This gives the following expression for the vertex function of the ideal chain $\Gamma_{pi,qj}^{(\text{ideal})}(\bm{q})$:
\begin{equation}
 \label{Gamma_ideal}
 \Gamma_{pi,qj}^{(\text{ideal})}(\bm{q}) = \frac{\delta_{pq}}{\bar{\phi}_{p}} g_{p,ij}(\bm{q}) 
\end{equation}
where $g_{p,ij}(\bm{q})$ is the inverse of the matrix $h_{p,ij}(\bm{q})$:
\begin{equation}
 \label{h_g_ortho}
 \sum_{j} h_{p,ij}(\bm{q}) g_{p,jk}(\bm{q}) = \delta_{ik}
\end{equation}

The effect of the interaction among the monomer units can be taken into account 
by adding the following interaction energy term
\begin{equation}
 \label{interaction_energy}
 F_{\text{int}} \left[\{\phi_{pi}(\bm{r})\}\right] 
   = \frac{1}{2} \sum_{pi,qj} \int d\bm{r} \, 
   \chi_{pi,qj} \phi_{pi}(\bm{r}) \phi_{qj}(\bm{r}) 
\end{equation}
where $\chi_{pi,qj}$ is the $\chi$ parameter for the interaction between 
the monomers in the block $(p,i)$ and those in the block $(q,j)$.  
Eqs \eqref{Gamma_ideal} and \eqref{interaction_energy} give the
following vertex function:
\begin{equation}
 \label{Gamma_int}
 \Gamma_{pi,qj}(\bm{q}) = \frac{\delta_{pq}}{\bar{\phi}_{p}} g_{p,ij}(\bm{q})
                        + \chi_{pi,qj}
\end{equation}

\subsection{Asymptotic Behavior of the Vertex Function for Large and Small
Wave Vector }

Though $\Gamma_{pi,qj}(\bm{q})$ can be obtained by the formula given above,
it is desirable to have an analytical expression for $\Gamma_{pi,qj}(\bm{q})$.
As it was first noticed by Leibler \cite{Leibler-1980}, 
the characteristics of $\Gamma_{pi,qj}(\bm{q})$ 
is that it diverges as $1/\bm{q}^2$
for small $\bm{q}$.  Ohta and Kawasaki utilized this fact in deriving 
their approximate density functional for the free energy of diblock copolymers. 
The divergence of $\Gamma_{pi,qj}(\bm{q})$ for small $\bm{q}$ reflects the
topological structure of the block copolymer.
The vertex function $\Gamma_{pi,qj}(\bm{q})$ also diverges 
for large $\bm{q}$ region in such a way as 
$\Gamma_{pi,qj}(\bm{q}) \propto \bm{q}^2$. Our strategy of constructing
the approximate free energy functional is to get an expression which 
correctly describe the 
asymptotic behavior of $\Gamma_{pi,qj}(\bm{q})$
for large $\bm{q}$ and small $\bm{q}$ region. Therefore we first 
study the behavior of $g_{p,ij}(\bm{q})$ 
for large $\bm{q}$ and small $\bm{q}$ region.

For large $\bm{q}$, the behavior of $g_{p,ij}(\bm{q})$ is easily seen.
In the limit of $\bm{q}^{2} \to \infty$, the
correlation function $h_{p,ij}(\bm{q})$ is totally determined by the
local structure of the polymer.  From eq~\eqref{correlation_ideal}, 
it follows
\begin{equation}
 \label{h_limit_infty}
 h_{p,ij}(\bm{q}) = \frac{12 \delta_{ij} f_{pi}}{b^{2}}\frac{1}{\bm{q}^{2}}  
                    + \dotsb, \qquad \bm{q}^{2} \to \infty
\end{equation}
Thus $g_{p,ij}$ is given by
\begin{equation}
 \label{g_limit_infty}
 g_{p,ij}(\bm{q}) = \frac{\delta_{ij} b^{2}}{12 f_{pi}} \bm{q}^{2} 
                    + \dotsb , \qquad \bm{q}^{2} \to \infty
\end{equation}

The behavior of $g_{p,ij}(\bm{q})$ for small $\bm{q}$ region is less obvious.
In the limit of $\bm{q}^{2} \to 0$,  eq~\eqref{correlation_ideal} can be
expanded into the power series of $\bm{q}^{2}$.
\begin{equation}
 \label{h_limit_0}
 h_{p,ij}(\bm{q}) = N_{p} f_{pi} f_{pj} - H_{p,ij} \bm{q}^{2} 
                       + \dotsb , \qquad \bm{q}^{2} \to 0
\end{equation}
where the expansion coefficient $H_{p,ij}$ is  given by
\begin{equation}
 \label{H_def}
 H_{p,ij} = 
  \begin{cases}
   \displaystyle \frac{1}{18} N_{p}^{2} f_{pi}^{3} b^{2} & (i = j) \\
   \displaystyle 
   N_{p} f_{pi} f_{pj} \left[ \frac{1}{12}  N_{p} ( f_{pi} + f_{pj} )b^{2}  + \frac{1}{6}l_{p,ij}^{2} \right]
   & (i \neq j)
  \end{cases}
\end{equation}
Eq~\eqref{h_limit_0} indicates that $\det (h_{p,ij}(\bm{q}))$
becomes zero at $\bm{q}^{2}=0$ (since $\det(N_pf_{pi}f_{pj})=0$).
Therefore the matrix equation \eqref{h_g_ortho} becomes singular 
for $\bm{q}^{2} \to 0$.
It is shown in the Appendix~\ref{detail_calculation} that the solution of eq~\eqref{h_g_ortho} for
small $\bm{q}$ can be written as
\begin{equation}
 \label{g_limit_0}
 g_{p,ij}(\bm{q}) =  \frac{A_{p,ij}}{\bm{q}^{2}} 
                        + \dotsb   \qquad \bm{q}^{2} \to 0
\end{equation}
where
\begin{equation}
 \label{Gi_def}
 A_{p,ij} = - \left(H_{p}^{-1}\right)_{ij} 
 + \frac{\displaystyle \sum_{kl} \left(H_{p}^{-1}\right)_{ik} f_{pk} f_{pl} \left(H_{p}^{-1}\right)_{lj}}{\displaystyle \sum_{kl} f_{pk} \left(H_{p}^{-1}\right)_{kl} f_{pl}}
\end{equation}
where $\left(H_{p}^{-1}\right)_{ij}$ is the $ij$ component of the
inverse matrix of $\left(H_{p}\right)_{ij}$:
\begin{equation}
 \label{inv_H}
 \sum_k H_{p,ij} \left(H_{p}^{-1}\right)_{jk} = \delta_{ik}
\end{equation}

\subsection{Approximate Expression for the Vertex Function}

Having seen the asymptotic behavior of $g_{p,ij}(\bm{q})$, 
we now seek an approximate
expression for $g_{p,ij}(\bm{q})$ which can be used in the entire
region of $\bm{q}$.  Considering the asymptotic behavior 
in the two limits of $\bm{q}^2 \to 0$ and $\bm{q}^2 \to \infty$, 
we use the following expression
\begin{equation}
 \label{g_approx_wavenumber}
   g_{p,ij}(\bm{q}) 
   =   \frac{A_{p,ij}}{\bm{q}^{2}}
     + C_{p,ij}
     + \frac{\delta_{ij} b^{2}}{12 f_{pi}} \bm{q}^{2}  
\end{equation}
where the constant $C_{p,ij}$ is chosen so that
eq~\eqref{g_approx_wavenumber}
gives a good approximation in the intermediate region.
We use the following procedure to determine $C_{p,ij}$.

For $i = j$, $g_{p,ij}(\bm{q})$ has a minimum at 
$\bm{\qstar}_{pi}^{2} = \sqrt{12 f_{pi} A_{p,ii} / b^{2}}$. 
We thus
determined $C_{p,ii}$ so that the minimum value agrees with the
exact value.  This gives
\begin{equation}
 \label{G_0_def_1}
 C_{p,ii} 
 = \left(h_{p}^{-1}(\bm{\qstar}_{pi})\right)_{ii}
 - \frac{A_{p,ii}}{\bm{\qstar}_{pi}^{2}}
 - \frac{b^{2}}{12 f_{pi}} \bm{\qstar}_{pi}^{2}
\end{equation}
where $\left(h_{p}^{-1}(\bm{\qstar}_{pi})\right)_{ij}$ stands for the 
$ij$ component of the inverse matrix $h_{p,ij}(\bm{q})$ at
$\bm{q}=\bm{\qstar}_{pi}$. 

For $i \ne j$, $g_{p,ij}(\bm{q})$ is a monotonically decreasing function 
of $\bm{q}^{2}$ and approaches to a constant value $C_{p,ij}$ for $\bm{q}^{2} \to \infty$.
We thus determined $C_{p,ij}$ by $(h_{p}^{-1}(\infty))_{ij}$.
This value is given by (see Appendix~\ref{detail_calculation})
\begin{equation}
 \label{G_0_def_2}
 C_{p,ij} = 
 \begin{cases}
  \displaystyle -\frac{1}{4 N_{p} f_{pi} f_{pj}} & (l_{p,ij}^{2} = 0) \\
  \displaystyle 0 & (l_{p,ij}^{2} \neq 0)
 \end{cases}
\end{equation}

From eqs \eqref{Gamma_int} and \eqref{g_approx_wavenumber}, 
the second order vertex function is obtained as
\begin{equation}
 \label{gamma_approx_wavenumber}
   \Gamma_{pi,qj}(\bm{q}) \approx \,
   \frac{\delta_{pq}}{\bar{\phi}_{p}} 
     \left[ 
           \frac{A_{p,ij}}{\bm{q}^{2}}
         + C_{p,ij} 
         + \frac{\delta_{ij} b^{2}}{12 f_{pi}} \bm{q}^{2} 
     \right]     
   + \chi_{pi,qj}
\end{equation}

This gives the following free energy in the real space representation:
\begin{equation}
 \label{freeenergy_weaksegregation}
 \begin{split}
  & F \left[\{\phi_{pi}(\bm{r})\}\right] = F \left[\{f_{pi} \bar{\phi}_{p}\}\right] \\
  & \qquad + \sum_{p,ij} \int d\bm{r} d\bm{r}' \, \frac{A_{p,ij}}{2 \bar{\phi}_{p}} 
  \mathcal{G}(\bm{r} - \bm{r}') \delta\phi_{pi}(\bm{r}) \delta\phi_{pj}(\bm{r}') \\
  & \qquad + \sum_{p,ij} \int d\bm{r} \, \frac{C_{p,ij}}{2 \bar{\phi}_{p}}
  \delta\phi_{pi}(\bm{r}) \delta\phi_{pj}(\bm{r}) \\
  & \qquad + \sum_{p,i} \int d\bm{r} \, \frac{b^{2}}{24 f_{pi} \bar{\phi}_{p}}
  \left| \nabla \delta\phi_{pi}(\bm{r}) \right|^{2} \\
  & \qquad + \sum_{pi,qj} \int d\bm{r} \, \frac{\chi_{pi,qj}}{2}
  \delta\phi_{pi}(\bm{r}) 
  \delta\phi_{qj}(\bm{r}) \\
  & \qquad + \dotsb
 \end{split}
\end{equation}
This free energy functional is valid for
$|\delta\phi_{pi}(\bm{r})| \ll 1$. 

\subsection{Test of the Approximate Free Energy for Small Perturbation}
We now test the accuracy of eq \eqref{freeenergy_weaksegregation} by 
calculating the density correlation function of
block copolymers. The density 
correlation function $S_{pi,qj}(\bm{q})$ can be calculated from $\Gamma_{pi,qj}(\bm{q})$
by solving
eq~\eqref{S_G_ortho}.  For polymer melts, an additional constraint
called the incompressible condition, is usually imposed.  
This condition is written as
\begin{equation}
 \label{incompressible_condition_real_wavenumber}
   \sum_{p,i} \phi_{pi}(\bm{r})= 1, 
   \quad \mbox{or}, \quad
   \sum_{p,i} \phi_{pi}(\bm{q})= 0 \qquad \text{for} \quad \bm{q} \neq 0
\end{equation}
This condition is equivalent to assume the following interaction energy
\begin{equation}
 \label{incompressible_condition_chi}
   \tilde {\chi} \sum_{pi,qj} \phi_{pi}(\bm{q})\phi_{qj}(-\bm{q})
\end{equation}
and taking the limit of $\tilde{\chi} \to \infty $.  This gives the
following density correlation function:
\begin{equation}
 \label{S_RPA}
   S_{pi,qj}(\bm{q}) =\left(\Gamma^{-1}\right)_{pi,qj}(\bm{q}) 
       - \frac{ \displaystyle
          \left[ \sum_{r,k}\left(\Gamma^{-1}\right)_{pi,rk}(\bm{q}) \right]
          \left[ \sum_{r,k}\left(\Gamma^{-1}\right)_{qj,rk}(\bm{q}) \right]
         }
         { \displaystyle
           \sum_{r,s,k,l}(\Gamma^{-1})_{rk,sl}(\bm{q})
         } 
\end{equation}
The second term comes from the incompressible condition.  

Figure~\ref{S_diblock} shows the scattering functions for AB diblock polymer
melts for $\chi_{AB}=0$. 
Here $S(\bm{q})$ is defined by
\begin{equation}
 \label{S_AB_def}
 S(\bm{q}) = S_{AA}(\bm{q}) + S_{BB}(\bm{q}) - 2 S_{AB}(\bm{q})
\end{equation}
The dashed line denote the exact scattering function calculated by
eqs~\eqref{S_G_ortho}, \eqref{Gamma_int} and \eqref{S_AB_def}.
The solid line denote the result of the approximate scattering function
calculated by eqs~\eqref{S_G_ortho}, 
\eqref{gamma_approx_wavenumber} and \eqref{S_AB_def}.  It is seen that the
agreement is quite good.

Figure~\ref{S_triblock} shows the scattering functions for 
symmetric ABA triblock polymer melts. The agreement is again quite good.

%
%

\section{General Free Energy Functional}

\subsection{Free Energy Variation for Phase Separated State}

Eq~\eqref{freeenergy_weaksegregation} can be used only for the
homogeneous state where $|\delta\phi_{pi}(\bm{r})| \ll 1$.
We now seek a general expression for the free energy which can be used for
the phase separated state. The expression has to reduce to 
eq~\eqref{freeenergy_weaksegregation} for the case of small deviation
from the homogeneous state. Clearly such generalized expression is
not unique.
The correct way to handle such systems is to evaluate the
correlations under inhomogeneous density profile, but it is quite difficult
to evaluate them analytically (in most cases, it must be done
numerically). 
In the following,
we propose a plausible form of the generalized 
expression based on physical argument instead, 
and test its validity for some typical cases.

First we note that, for phase separated state, 
eq~\eqref{freeenergy_weaksegregation} is not valid
even if the density deviation $\delta\phi_{pi}(\bm{r})$ from
the equilibrium state is small: for macroscopically 
phase separated system, $\bar{\phi}_p$ 
in eq~\eqref{freeenergy_weaksegregation} should not be the
average of $\phi_p(\bm{r})$ for the entire volume; it should be
the average of $\phi_p(\bm{r})$ in the local region of the 
phase separated state where the point 
$\bm{r}$ and $\bm{r'}$ are located.  This consideration 
suggests the following
replacement
\begin{equation} 
  \bar{\phi}_p  \to
    \left[ \phi_{p}(\bm{r})\phi_{p}(\bm{r}') \right ]^{1/2}
\end{equation}
However such replacement does not give an analytically tractable form for the
free energy.  As an alternative,we used the following replacement:
\begin{equation}
 \label{phi_replacement}
  \bar{\phi}_p  \to 
    \left[ 
        \frac{\phi_{pi}(\bm{r})\phi_{pj}(\bm{r}')}{f_{pi}f_{pj}}
     \right ]^{1/2}
\end{equation}
This gives the following free energy functional for small variation of
$\phi_{pi}(\bm{r})$.
\begin{equation}
 \label{freeenergy_weaksegregation_1}
  \begin{split}
   & \delta^{(2)} F \left[ \{ \phi_{pi}(\bm{r}) \} \right] = \\
   & \qquad \sum_{p,ij} \int d\bm{r} d\bm{r}' \, \frac{1}{2} \sqrt{f_{pi} f_{pj}} A_{p,ij}
   \mathcal{G}(\bm{r} - \bm{r}')
   \frac{\delta\phi_{pi}(\bm{r})}{\sqrt{\phi_{pi}(\bm{r})}}
   \frac{\delta\phi_{pj}(\bm{r}')}{\sqrt{\phi_{pj}(\bm{r}')}} \\
   & \qquad + \sum_{p,ij} \int d\bm{r} \, \frac{1}{2} \sqrt{f_{pi} f_{pj}} C_{p,ij}
   \frac{\delta\phi_{pi}(\bm{r}) }{\sqrt{\phi_{pi}(\bm{r})}}
   \frac{\delta\phi_{pj}(\bm{r})}{\sqrt{\phi_{pj}(\bm{r})}} \\  
   & \qquad + \sum_{pi} \int d\bm{r} \, \frac{b^{2}}{24}
   \frac{\left| \nabla \delta\phi_{pi}(\bm{r})\right|^{2}}{{\phi_{pi}(\bm{r})}} \\
   & \qquad + \sum_{pi,qj} \int d\bm{r} \, \frac{\chi_{pi,qj}}{2}
   \delta\phi_{pi}(\bm{r}) \delta\phi_{qj}(\bm{r}) \\
   & \qquad + \dotsb
  \end{split}
\end{equation}

\subsection{Free Energy for the General Case}
The right hand side of eq.\eqref{freeenergy_weaksegregation_1} represents
the second order functional variation of the functional  
$F \left[\{\phi_{pi}(\bm{r})\}\right]$ with respect to $\phi_{pi}(\bm{r})$.
The equation is then regarded as a second order functional differential equation.
From the equation, the functional 
$F \left[\{\phi_{pi}(\bm{r})\}\right]$ 
is uniquely determined.  The result is
\begin{equation}
 \label{freeenergy_strongsegregation}
 \begin{split}
  & F \left[\{\phi_{pi}(\bm{r})\}\right] = \\
  & \qquad \sum_{p,ij}  \int d\bm{r} d\bm{r}' \, 2 \sqrt{f_{pi} f_{pj}} A_{p,ij}  
  \mathcal{G}(\bm{r} - \bm{r}') \sqrt{\phi_{pi}(\bm{r}) \phi_{pj}(\bm{r}')} \\
  & \qquad + \sum_{pi}  \int d\bm{r} \, f_{pi} C_{p,ii} 
  \phi_{pi}(\bm{r}) \ln \phi_{pi}(\bm{r}) \\
  & \qquad + \sum_{p,i \neq j} \int d\bm{r} \, 2 \sqrt{f_{pi} f_{pj}} C_{p,ij} 
  \sqrt{\phi_{pi}(\bm{r}) \phi_{pj}(\bm{r})} \\
  & \qquad + \sum_{pi} \int d\bm{r} \, \frac{b^{2}}{24 \phi_{pi}(\bm{r})}
  \left| \nabla \phi_{pi}(\bm{r}) \right|^{2} \\
  & \qquad + \sum_{pi,qj} \int d\bm{r} \, \frac{\chi_{pi,qj}}{2} \phi_{pi}(\bm{r}) \phi_{qj}(\bm{r})
 \end{split}
\end{equation}
It is straightforward to check that the second order functional 
variation of eq~\eqref{freeenergy_weaksegregation_1}
with respect to $\delta\phi_{pi}(\bm{r})$ leads eq~\eqref{freeenergy_weaksegregation}.
This free energy functional does
not depend on $\bar{\phi}_{p}$ and can be used for strong segregation.

Of course there ane many other possible free energy functional which
reduces to eq~\eqref{freeenergy_weaksegregation}, but
eq~\eqref{freeenergy_strongsegregation} has several advantages.
\begin{enumerate}
 \item For homopolymer blends, 
       eq~\eqref{freeenergy_strongsegregation} gives the Flory-Huggins-de Gennes
       type \cite{Joanny-Leibler-1978,deGennes-1980} (or Lifshitz
       type \cite{Grosberg-Khokhlov-book}) free energy.  This is shown in
       section \ref{homopolymer_blends}.
 \item For the ordered phase of diblock copolymers,
       eq~\eqref{freeenergy_strongsegregation} gives the results equivalent to
       that of the
       Ohta-Kawasaki theory \cite{Ohta-Kawasaki-1986,Ohta-Kawasaki-1990}.
       This will be demonstrated in section \ref{diblock_copolymer_melts}.
 \item The minimization of eq~\eqref{freeenergy_strongsegregation} with respect
       to $\phi_{pi}(\bm{r})$ usually requires numerical calculation for which
       eq~\eqref{freeenergy_strongsegregation} has an advantage. 
       We found that by using the variable  
       $\psi_{pi}(\bm{r}) \equiv \sqrt{\phi_{pi}(\bm{r})} $ rather than 
       $\phi_{pi}(\bm{r}) $ itself, we can improve the numerical stability
       significantly. The detail of the numerical procedure is described in
       Appendix~\ref{numerical_scheme}.
\end{enumerate}


%
%

\section{Test of the Free Energy}
\label{specialcase}
\label{validation}

In this section we apply the free energy expression \eqref{freeenergy_final}
to special cases, and discuss its accuracy.

\subsection{Homopolymer Blends}
\label{homopolymer_blends}
For homopolymer, $A_{p}$ becomes identically equal to zero since eq
\eqref{Gi_def} gives
\begin{equation}
 A_{p} = - N_{p}^{-1} + \frac{N_{p}^{-1} N_{p}^{-1}}{N_{p}^{-1}} = 0
\end{equation}
and there is no long range interaction.

$C_{p}$ can be calculated as follows.
\begin{equation}
   C_{p} =\left( h_{p}(0) \right )^{-1} = \frac{1}{N_{p}}
\end{equation}

From eq~\eqref{freeenergy_strongsegregation}, the free energy functional
for the system can be described as
\begin{equation}
 \label{freeenergy_homopolymer}
 \begin{split}
  & F \left[\{\phi_{p}(\bm{r})\}\right] = \\
  & \qquad \sum_{p} \int d\bm{r} \,  \frac{1}{N_{p}} \phi_{p}(\bm{r}) \ln \phi_{p}(\bm{r}) \\
  & \qquad + \sum_{p} \int d\bm{r} \, \frac{b^{2}}{24 \phi_{p}(\bm{r})} \left| \nabla \phi_{p}(\bm{r}) \right|^{2} \\
  & \qquad + \sum_{p,q} \int d\bm{r} \, \frac{\chi_{pq}}{2} \phi_{p}(\bm{r}) \phi_{q}(\bm{r})
 \end{split}
\end{equation}
Thus our free energy functional reduces to the Flory-Huggins-de
Gennes type free energy functional
\cite{deGennes-1980} (strictly speaking, the free energy given in
ref~\cite{deGennes-1980} has the factor $1/36$ in the second term of
eq~\eqref{freeenergy_homopolymer}. This is because in ref~\cite{deGennes-1980},
the factor is determined from the behavior of $h(\bm{q})$ for small 
$\bm{q}$ region, while we determined it using the behavior for large 
$\bm{q}$ region. The factor $1/24$ agree with that
given by Lifshitz et. al. \cite{Grosberg-Khokhlov-book}).

\subsection{Diblock Copolymer Melts}
\label{diblock_copolymer_melts}
For diblock copolymer melts, $A_{ij}$ and $C_{ij}$ are given by
\begin{align}
 A_{ij} & = \frac{9}{N^{2} b^{2} f^{2} (1 - f)^{2}}
 \begin{bmatrix}
  (1 - f)^{2} & - f (1 - f) \\
  - f (1 - f) & f^{2}
 \end{bmatrix} \\
 C_{ij} & = \frac{1}{N f (1 - f)}
 \begin{bmatrix}
  s(f) & -1 / 4 \\
  -1 / 4 & s(1 - f)
 \end{bmatrix}
\end{align}
where $b_{A} = b_{B} = b, f_{A} = 1- f_{B} = f$ and $s(f)$ is the
function which is determined from eqs~\eqref{G_0_def_1}, \eqref{G_0_def_2}.

From eq~\eqref{freeenergy_weaksegregation} the free energy functional
for the weak segregation limit will be
\begin{equation}
 \begin{split}
  & F \left[\phi(\bm{r})\right] = \\
  & \qquad \int d\bm{r} d\bm{r}' \, \frac{9}{2 N^{2} b^{2} f^{2} (1 - f)^{2}} \mathcal{G}(\bm{r} - \bm{r}') \delta\phi(\bm{r}) \delta\phi(\bm{r}') \\
  & \qquad - \int d\bm{r} \, \bar{\chi} \delta\phi^{2}(\bm{r}) \\
  & \qquad + \int d\bm{r} \, \frac{b^{2}}{24 f (1 - f)} \left| \nabla \delta\phi(\bm{r}) \right|^{2} \\
  & \qquad + \mathcal{W}\left[\delta\phi(\bm{r})\right]
 \end{split}
\end{equation}
where we set $\phi(\bm{r}) = \phi_{A}(\bm{r}) = 1 - \phi_{B}(\bm{r})$,
$F\left[\bar{\phi}\right] = 0$ and $\bar{\chi} = \chi_{AB} - [s(f) + s(1
- f) + 1 / 2] / 2 N f (1 - f)$. $\mathcal{W}\left[\delta\phi(\bm{r})\right]$
is the contribution of the higher order terms
for $\delta\phi(\bm{r})$.
In this case the free energy functional reduces to the
Ohta-Kawasaki type free energy functional \cite{Ohta-Kawasaki-1986}.

For strong segregation, we get the following free energy functional from
eq~\eqref{freeenergy_strongsegregation}.
\begin{equation}
 \label{freeenergy_diblock_strong}
 \begin{split}
  & F \left[\phi(\bm{r})\right] = \\
  & \qquad \int d\bm{r} d\bm{r}' \,\frac{18}{N^{2} b^{2} f (1 - f)} \mathcal{G}(\bm{r} - \bm{r}') \tau(\bm{r}) \tau(\bm{r}')\\
  & \qquad + \int d\bm{r} \, \frac{1}{N} \left[ \frac{s(f)}{1 - f} \phi(\bm{r}) \ln \phi(\bm{r}) + \frac{s(1 - f)}{f} \left(1 - \phi (\bm{r}) \right) \ln \left(1 - \phi(\bm{r})\right)\right] \\
  & \qquad - \int d\bm{r} \, \frac{1}{N \sqrt{f (1 - f)}} \sqrt{\phi(\bm{r}) \left(1 - \phi(\bm{r}) \right)} \\
  & \qquad + \int d\bm{r} \, \frac{b^{2}}{24 \phi(\bm{r}) \left(1 - \phi(\bm{r})\right)} \left| \nabla \phi(\bm{r}) \right|^{2} \\
  & \qquad + \int d\bm{r} \, \chi \phi(\bm{r}) \left( 1 - \phi(\bm{r}) \right)
 \end{split}
\end{equation}
where $\tau(\bm{r}) \equiv \sqrt{(1 - f) \phi(\bm{r})} - \sqrt{f \left(1
- \phi(\bm{r}) \right)}$ and $\chi = \chi_{AB}$. The interfacial energy
in eq~\eqref{freeenergy_diblock_strong} (the fourth term) does not depend
on $f$.
This is consistent with the improved form of the Ohta-Kawasaki theory \cite{Ohta-Kawasaki-1990}.

Figure~\ref{AB_phasediagram} shows the phase diagram for AB diblock
copolymer melts.
The solid line is the results of the DF theory and
the dashed line is the result of Matsen Bates \cite{Matsen-Bates-1996}
which is based on the SCF theory.  The critical point
predicted by our DF is $\chi N = 10.553778$, while that predicted by the
SCF theory is $\chi N= 10.495$. The results of the DF and the SCF do not agree
quantitatively but they agree well qualitatively. Especially
notice that the DF gives a stable double gyroid phase between the lamellar
and the hexagonal cylinder phase for small $\chi N$. It is difficult to
get such a result by the previous density functional theories (the
Leibler theory or the Ohta-Kawasaki theory).

Figures~\ref{AB_period} shows the equilibrium periods for AB diblock
copolymer melts plotted against $\chi N$. 
Our DF theory does not agree well with the SCF theory, but it 
reproduces the scaling feature correctly: $D
\propto N^{1/2}$ for weak segregation and $D \propto N^{3/2}$ for strong
segregation. The numerical disagreement is the nature of the Ohta-Kawasaki type
approximation and arises from the approximation for the
vertex functions (eq~\eqref{gamma_approx_wavenumber}).

Figures~\ref{AB_equilibrium} shows the equilibrium
density profile of AB diblock copolymer melts.
The solid line is the results of the DF theory and
the dashed line is the result of the SCF theory\cite{Matsen-Bates-1996}.
The result of the DF  theory deviates significantly from that of  SCF theory for weak
segregation, while they become close to each other 
for strong segregation.

\subsection{A,B Homopolymer / AB Diblock Copolymer Blends}
We applied eq~\eqref{freeenergy_final} to A,B homopolymer / AB diblock
copolymer blends.
The free energy model for A,B homopolymer / AB diblock copolymer melts
was proposed and studied by Kawakatsu
\cite{Kawakatsu-1994}. Our free energy functional has similar form to
the Kawakatsu's theory.

Figure~\ref{A_B_AB_equilibrium} is equilibrium structures for an A,B
homopolymer / AB diblock copolymer blend. We set
$\bar{\phi}_{A} = \bar{\phi}_{B} = 0.475, \bar{\phi}_{AB} = 0.05, 
N_{A} = N_{B} = N_{AB} = 40, f_{AB,A} = f_{AB,B} = 0.5, \chi_{AB} = 1$.
The result of the SCF theory was obtained by
using the SUSHI engine in OCTA system \cite{SUSHI-manual}.
Again the agreement between the DF theory and the SCF theory is not
perfect, but reasonable considering that no adjustable parameter is used
in these comparison.

\subsection{AB diblock coplymer / C homopolymer blends}
As the last example, we show the results of AB diblock copolymer / C
homopolymer blends \cite{Ohta-Ito-1995}. These blends cause the macrophase separation as
well as the microphase separation and show various interesting
structures. Figures~\ref{AB_C_equilibrium1} and \ref{AB_C_equilibrium2}
are equilibrium structures for AB diblock copolymer / C homopolymer
blends (the parameters are
$N_{AB} = 10, N_{C} = 20, f_{AB,A} = f_{AB,B} = 0.5,
\bar{\phi}_{AB} = 0.2, \bar{\phi}_{C} = 0.8,
\chi_{AB} = 1.2, \chi_{BC} = 1, \chi_{CA} = 0.5$ and
$N_{AB} = 10, N_{C} = 20, f_{AB,A} = 0.35, f_{AB,B} = 0.65,
\bar{\phi}_{AB} = 0.3, \bar{\phi}_{C} = 0.7,
\chi_{AB} = 1.75, \chi_{BC} = 1, \chi_{CA} = 0.5$, respectively).
The simulation has been done for the three dimensional periodic system of
$64 \times 64 \times 64$ lattice points (the system size is $40 b
\times 40 b \times 40 b$).
The structure of the AB diblock copolymer in
Figure~\ref{AB_C_equilibrium1} is considered to be the ``onion structure''
observed by the experiments \cite{Koizumi-Hasegawa-Hashimoto-1994}.
It is also noted that each simulation starts from the homogeneous state, and
gives the results shown there after about 4 hours on the
3.0GHz Xeon workstation. It is quite difficult to get
these results with such short computational time by the SCF simulation.

\subsection{Comparison with the other density functional theory}
In this section, we compare our theory with the density functional theory proposed
by Bohbot-Raviv and Wang for block copolymer melts \cite{BohbotRaviv-Wang-2000}. 
They used the Flory-Huggins free energy for the disconnected blocks as the
reference and took into account of the effect of chain correlation by the second order 
vertex function with the RPA: their free energy functional is expressed as
\begin{align}
 \label{freeenergy_bw}
 F \left[\{\phi_{i}(\bm{r})\}\right] = &
 F_{\text{ref}} \left[\{\phi_{i}(\bm{r})\}\right] 
 + \Delta F \left[\{\Delta\phi_{i}(\bm{q})\}\right] \\
 F_{\text{ref}} \left[\{\phi_{i}(\bm{r})\}\right] = &
 \sum_{i} \int d\bm{r} \, \frac{1}{f_{i} N} \phi_{i}(\bm{r}) \ln \phi_{i}(\bm{r}) \\
 \begin{split}
  \Delta F \left[\{\Delta\phi_{i}(\bm{q})\}\right] = &
  \frac{1}{2} \sum_{i,j} \frac{1}{(2 \pi)^{d}}\int d\bm{q} \,
  \Gamma_{i,j}(\bm{q}) \Delta\phi_{i}(\bm{q})
  \Delta\phi_{j}(-\bm{q}) \\
  & - \frac{1}{2} \sum_{i} \frac{1}{(2 \pi)^{d}} \int d\bm{q} \frac{1}{f_{i} N} \Delta\phi_{i}(\bm{q}) \Delta\phi_{i}(-\bm{q})  
 \end{split}
\end{align}
where
\begin{equation}
 \Delta\phi_{i}(\bm{r}) = \phi_{i}(\bm{r}) - f_{i}
\end{equation}
Eq~\eqref{freeenergy_bw} is exactly same as eq~\eqref{intro_1} except for higher
order terms for $\Delta\phi_{i}(\bm{r})$ (it can be easily shown by setting $\bar{\phi} = 1$ and expanding
eq~\eqref{freeenergy_bw} by series of $\Delta\phi_{i}(\bm{r})$).

Though their approach is simple and straightforward their theory cannot
overcome one well-known difficulty.
Ohta and Kawasaki noticed that the block ratio dependence of the
interfacial tension calculated from
the RPA is inconsistent with other theories \cite{Ohta-Kawasaki-1990}.
They then proposed to replace $f$, the block ratio, by the local density
$\phi(\bm{r})$. Without this correction, the free energy
eq~\eqref{freeenergy_bw} cannot give correct block ratio dependence.
The replacement eq~\eqref{phi_replacement} is essentially same as the
replacement by Ohta and Kawasaki and therefore our free
energy~\eqref{freeenergy_final} does not have such difficulty.

%
%

\section{Conclusion}
\label{conclusion}

We have given an expression of the free energy of blends of block copolymers
expressed as a functional of the density distribution of monomers in each block.
All parameters in the expression are determined by the polymer structure parameter
(such as the degree of polymerization, the branching structure, and block ratio
etc.) and the $\chi$ parameter.  
For homopolymer blends, the expression is shown to reduce to the
Flory-Huggins-de Gennes theory.  For diblock copolymer melts, the expression
gives results equivalent to the Ohta-Kawasaki theory.
Thus the expression is expected to work in more general case. 
We have applied the expression for several cases, and found reasonable
agreement with the SCF calculation. Of course, these examples do not 
guarantee the validity of our approximation in the general case, and more
work is clearly needed to test the validity of our expression.

The present theory is most useful by combining it to the SCF calculation.
It uses exactly the same parameter set as the SCF theory (this is
in contrast to the other mesoscopic approaches such as the dissipative
particle dynamics (DPD) \cite{Groot-Warren-1997}). Therefore
the theory can be used to get the initial structure for the SCF calculation,
or it can be used to get an overview of the phase diagram of block copolymers.
These applications will be pursued in future.


%
%

\section*{Acknowledgment}
The authors thank Prof. Takao Ohta (Kyoto University) and
Prof. Toshihiro Kawakatsu (Tohoku University) for encouraging and useful
discussions. The present work is supported by 
the Japan Science and Technology Agency (CREST-JST).


\section*{Appendix}
\appendix
%
%

\section{Calculation of $A_{p,ij}$ and $C_{p,ij}$}
\label{detail_calculation}

\subsection{Calculation of $A_{p,ij}$}

In this section we derive eq~\eqref{Gi_def}.
At the limit of $\bm{q}^{2} \to 0$, $h_{p,ij}(\bm{q})$ can be expanded
as follows.
\begin{equation}
 \label{h_limit_0_detail}
 h_{p,ij}(\bm{q}) = N_{p} f_{pi} f_{pj} - H_{p,ij} \bm{q}^{2} + \dotsb
\end{equation}
where $H_{p,ij}$ is given by eq~\eqref{H_def}.
As we are considering one ideal chain,
we omit the subscript ``$p$'' and describe eq~\eqref{h_limit_0_detail}
in a matrix form.
\begin{equation}
 \bm{h}(\bm{q}) = N \bm{f} \transpose{\bm{f}} - \bm{H} \bm{q}^{2} + \dotsb
\end{equation}
where $\transpose{\bm{f}}$ is the transposed vector of $\bm{f}$.

We obtain $\bm{g}(\bm{q})$, the inverse matrix of $\bm{h}(\bm{q})$, in
the following form:
\begin{equation}
 \bm{g}(\bm{q}) = \bm{A} \frac{1}{\bm{q}^{2}} + \bm{B} + \dotsb
\end{equation}
From eq~\eqref{h_g_ortho},
\begin{equation}
  \bm{g}(\bm{q}) \cdot \bm{h}(\bm{q}) = \left( \bm{A} \frac{1}{\bm{q}^{2}} + \bm{B} + \dotsb \right) \cdot \left( N \bm{f} \transpose{\bm{f}} - \bm{H} \bm{q}^{2} + \dotsb \right) = \bm{E}
\end{equation}
and we get the following set of equations.
\begin{align}
\label{h_g_ortho_0}
& \bm{A} \cdot \bm{f} = 0 \\
\label{h_g_ortho_1}
& - \bm{A} \cdot \bm{H} + N \bm{B} \cdot  \bm{f} \transpose{\bm{f}} = \bm{E}
\end{align}
where $\bm{E}$ is unit matrix. Notice that the matrix $\bm{h}(\bm{q})$ is symmetric
and therefore the matrices $\bm{A}$ and $\bm{B}$ are also symmetric.
Multiplying $\bm{H}^{-1} \cdot \bm{f} \transpose{\bm{f}}$ to eq~\eqref{h_g_ortho_1} from the left
side and using the relation 
$\transpose{\bm{f}} \cdot \bm{A}=\transpose{(\bm{A} \cdot \bm{f})} =0$
which follows from eq~\eqref{h_g_ortho_0}, we get
\begin{equation}
 \label{h_g_ortho_1_mod}
  N \bm{B} \cdot \bm{f} \transpose{\bm{f}} \cdot \bm{H}^{-1} \cdot \bm{f} \transpose{\bm{f}} 
  = \bm{H}^{-1} \cdot \bm{f} \transpose{\bm{f}}
\end{equation}
Since $\transpose{\bm{f}} \cdot \bm{H}^{-1} \cdot \bm{f}$ is a scalar,
eq~\eqref{h_g_ortho_1_mod} gives
\begin{equation}
 \label{h_g_ortho_1_mod2}
  N \bm{B} \cdot \bm{f} \transpose{\bm{f}} = \frac{\bm{H}^{-1} \cdot \bm{f} \transpose{\bm{f}}}{ \transpose{\bm{f}} \cdot \bm{H}^{-1} \cdot \bm{f}}
\end{equation}
From eqs~\eqref{h_g_ortho_1} and \eqref{h_g_ortho_1_mod2} we have
\begin{equation}
 \bm{A} = 
   - \bm{H}^{-1} 
   + \frac{\bm{H}^{-1}  \cdot \bm{f} \transpose{\bm{f}} \cdot \bm{H}^{-1}}
        {\transpose{\bm{f}} \cdot  \bm{H}^{-1} \cdot \bm{f}}
\end{equation}
This gives eq~\eqref{Gi_def}.

\subsection{Calculation of $C_{p,ij}$}

In this section we calculate the asymptotic value of $g_{p,ij}(\bm{q})$
at $\bm{q}^{2} \to \infty$ for $i \neq j$.
%
%
From eq~\eqref{correlation_ideal}, $h_{p,ij}(\bm{q})$ can be described as
follows in matrix representation.
\begin{equation}
 \bm{h}(\bm{q}) = \bm{G} \frac{1}{\bm{q}^{2}} + \bm{K} \frac{1}{\bm{q}^{4}} + \dotsb
\end{equation}
where
\begin{align}
 \bm{G} = & \frac{12 f_{i} \delta_{ij}}{b^{2}}\\
 \bm{K} = & \begin{cases}
	     \displaystyle -\frac{72}{N b^{4}} & (i = j) \\
	     \displaystyle \frac{36}{N b^{4}} & (i \neq j, l_{ij}^{2} = 0) \\
	     0 & (i \neq j, l_{ij}^{2} \neq 0)
            \end{cases}
\end{align}
$g_{p,ij}(\bm{q})$ can be expressed as power series of $\bm{q}^{2}$.
\begin{equation}
 \bm{g}(\bm{q}) = \bm{B} \bm{q}^{2} + \bm{C} + \dotsb
\end{equation}
The equations to determine coefficients are
\begin{align}
& \bm{B} \bm{G} = \bm{E} \\
\label{h_g_ortho_2}
& \bm{B} \bm{K} + \bm{C} \bm{G}  = 0
\end{align}
Thus we get
\begin{equation}
 \begin{split}
  \bm{C} & = - \bm{B} \bm{K} \bm{G}^{-1} = - \bm{G}^{-1} \bm{K} \bm{G}^{-1} \\
  & = \begin{cases}
       \displaystyle \frac{1}{2 f_{i}^{2} N} & (i = j) \\
       \displaystyle - \frac{1}{4 f_{i} f_{j} N} & (i \neq j, l_{ij}^{2} = 0) \\
       0 & (i \neq j, l_{ij}^{2} \neq 0)
      \end{cases}
 \end{split}
\end{equation}
This gives eq~\eqref{G_0_def_2}.

\section{Numerical Scheme}
\label{numerical_scheme}

The minimization of eq~\eqref{freeenergy_strongsegregation} with respect
to $\phi_{pi}(\bm{r})$ usually requires the numerical calculation. Here
we briefly describe the numerical scheme we employed.

For numerical calculation, it is
convenient to introduce a
new order parameter $\psi_{pi}(\bm{r})$.
\begin{equation}
   \psi_{pi}(\bm{r}) \equiv \sqrt{\phi_{pi}(\bm{r})}
\end{equation}
For this order parameter, the free energy functional is written as 
\begin{equation}
 \label{freeenergy_final}
 \begin{split}
  & F \left[\{\psi_{pi}(\bm{r})\}\right] = \\
  & \qquad \sum_{p,ij} \int d\bm{r} d\bm{r}' \, 2 \sqrt{f_{pi} f_{pj}} A_{p,ij} \mathcal{G}(\bm{r} - \bm{r}') \psi_{pi}(\bm{r}) \psi_{pj}(\bm{r}') \\
  & \qquad + \sum_{pi} \int d\bm{r} \, 2 f_{pi} C_{p,ii} \psi_{pi}^{2}(\bm{r}) \ln \psi_{pi}(\bm{r}) \\
  & \qquad + \sum_{p,i \neq j} \int d\bm{r} \, 2 \sqrt{f_{pi} f_{pj}} C_{p,ij} \psi_{pi}(\bm{r}) \psi_{pj}(\bm{r}) \\
  & \qquad + \sum_{pi} \int d\bm{r} \, \frac{b^{2}}{6} \left| \nabla \psi_{pi}(\bm{r}) \right|^{2} \\
  & \qquad + \sum_{pi,qj} \int d\bm{r} \, \frac{\chi_{pi,qj}}{2} \psi_{pi}^{2}(\bm{r}) \psi_{qj}^{2}(\bm{r})
 \end{split}
\end{equation}
This expression is advantageous for numerical calculation as it does not have
the numerical instability associated with the logarithmic term of
eq~\eqref{freeenergy_strongsegregation}. 
The fourth term in eq~\eqref{freeenergy_final} is in agreement with that
proposed by Lifshitz \cite{Lifshitz-1969,Grosberg-Khokhlov-book}.

The equilibrium structure is obtained by minimizing the free energy
\eqref{freeenergy_final} under the following constraints
\begin{align}
 \label{incompressible_condition}
 \sum_{pi} \psi_{pi}^{2}(\bm{r}) & = 1 \\
 \label{conservation_condition}
 \int d\bm{r} \psi_{pi}^{2}(\bm{r}) & = V f_{pi} \bar{\phi}_{p}
\end{align}
The first equation represents the incompressible condition, and the
second equation represents the mass conservation for each block.
To take into account of these constraints
we use the Lagrangian multiplier method, and
added the following terms to eq \eqref{freeenergy_final}.
\begin{equation}
 \label{freeenergy_constraints}
 F_{\text{constraint}} \left[\{\psi_{pi}(\bm{r})\}\right] = 
 \sum_{pi} \int d\bm{r} \, \frac{1}{2} \left[ \lambda_{pi} + \kappa(\bm{r}) \right] 
                             \left(\psi_{pi}^{2}(\bm{r}) - f_{pi} \bar{\phi}_{p} \right)
\end{equation}
where $\lambda_{pi}$ and $\kappa(\bm{r})$ are the Lagrangian multipliers.

We used simple relaxation method to get equilibrium structures.
\begin{equation}
 \label{relaxation_method}
 \psi^{(n + 1)}_{pi}(\bm{r}) = \psi^{(n)}_{pi}(\bm{r}) - \omega \mu^{(n)}_{pi}(\bm{r})
\end{equation}
where the superscript $(n)$ means the number of the relaxation step, $\omega$
is a sufficiently small constant and $\mu_{pi}(\bm{r})$ is the chemical potential defined
as follows.
\begin{equation}
 \label{mu_def}
 \begin{split}
  \mu_{pi}(\bm{r}) \equiv & \frac{\delta F \left[\{\psi_{pi}(\bm{r})\}\right]}{\delta \psi_{pi}(\bm{r})}  + \frac{\delta F_{\text{constraint}} \left[\{\psi_{pi}(\bm{r})\}\right]}{\delta \psi_{pi}(\bm{r})} \\
  = & \sum_{j} 4 \sqrt{f_{pi} f_{pj}} A_{p,ij}  \int d\bm{r}' \mathcal{G}(\bm{r} - \bm{r}') \psi_{pj}(\bm{r}') \\
  & + 2 f_{pi} C_{p,ii} \left[ 2 \psi_{pi}(\bm{r}) \ln \psi_{pi}(\bm{r}) + \psi_{pi}(\bm{r}) \right] \\
  & + \sum_{j (j \neq i)} 4 \sqrt{f_{pi} f_{pj}} C_{p,ij} \psi_{pj}(\bm{r}) \\
  & - \frac{b^{2}}{3} \nabla^{2} \psi_{pi}(\bm{r}) \\
  & + \sum_{qj} 2 \chi_{pi,qj} \psi_{pi}(\bm{r}) \psi_{qj}^{2}(\bm{r}) \\
  & + [\lambda_{pi} + \kappa(\bm{r})] \psi_{pi}(\bm{r})
 \end{split}
\end{equation}
The terms including $\mathcal{G}(\bm{r} - \bm{r}')$ or $\nabla^{2}$ are
calculated by using the fast Fourier transform (the FFTW library \cite{Frigo-Johnson-1998} is used), and
other terms are calculated in the real space. It is noted that
eq~\eqref{mu_def} has no singularity at $\psi_{pi}(\bm{r}) = 0$ while
$\delta F \left[\{\phi_{pi}(\bm{r})\}\right] / \delta
\phi_{pi}(\bm{r})$, the chemical potential for $\phi_{pi}(\bm{r})$, is
singular at $\phi_{pi}(\bm{r}) = 0$.
Thus using the order parameter $\psi_{pi}(\bm{r})$ rather that
$\phi_{pi}(\bm{r})$ improves the numerical stability significantly.

The numerical method we employed in this work is not so efficient
We employed this method for simplicity,
but the algorithm need to be improved (for example, the ICCG method
will improve the convergence).




\clearpage

\begin{figure}[htbp]
 \centering
 \includegraphics{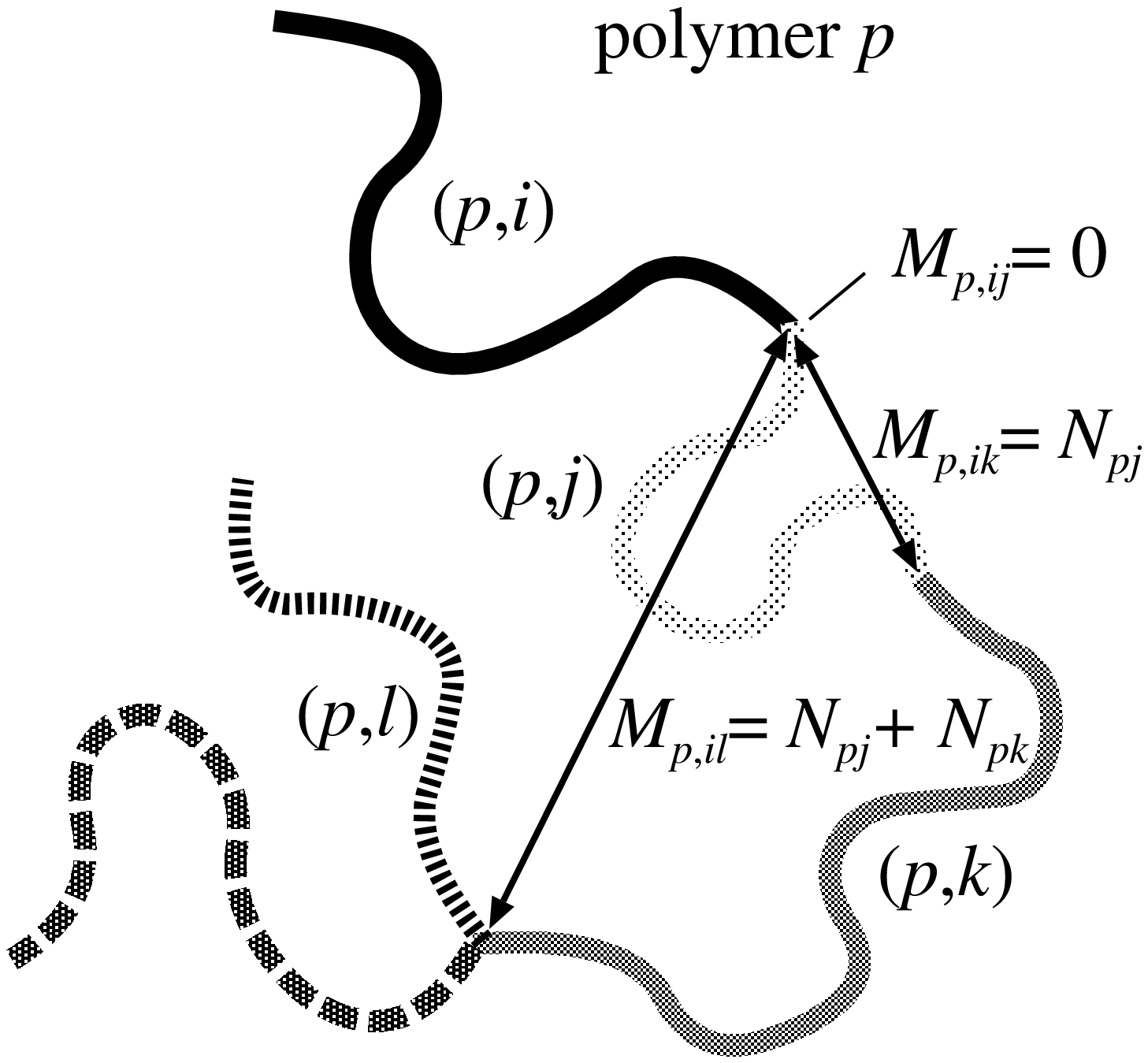}
 \caption{}
 \label{mean_square_distance}
\end{figure}

\begin{figure}[htbp]
 \centering
 \includegraphics{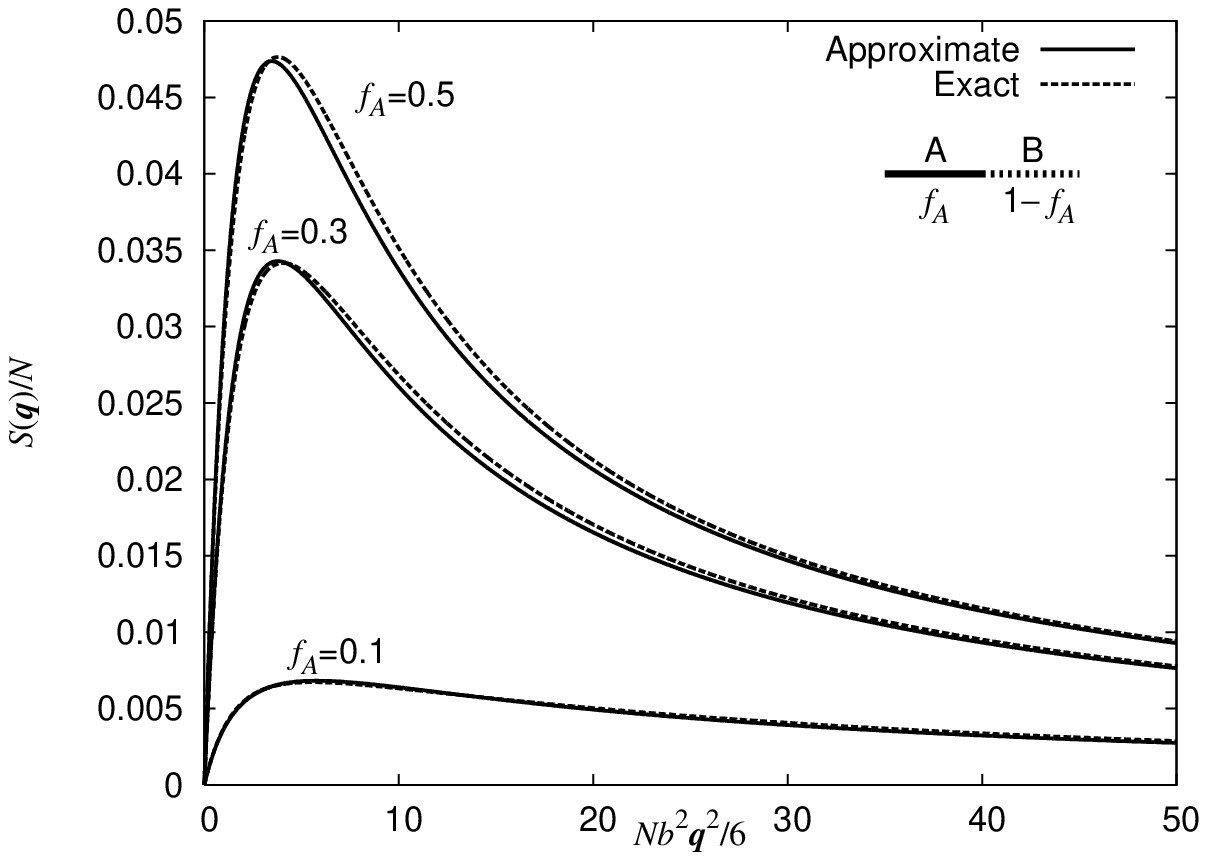}
 \caption{}
 \label{S_diblock}
\end{figure}

\begin{figure}[htbp]
 \centering
 \includegraphics{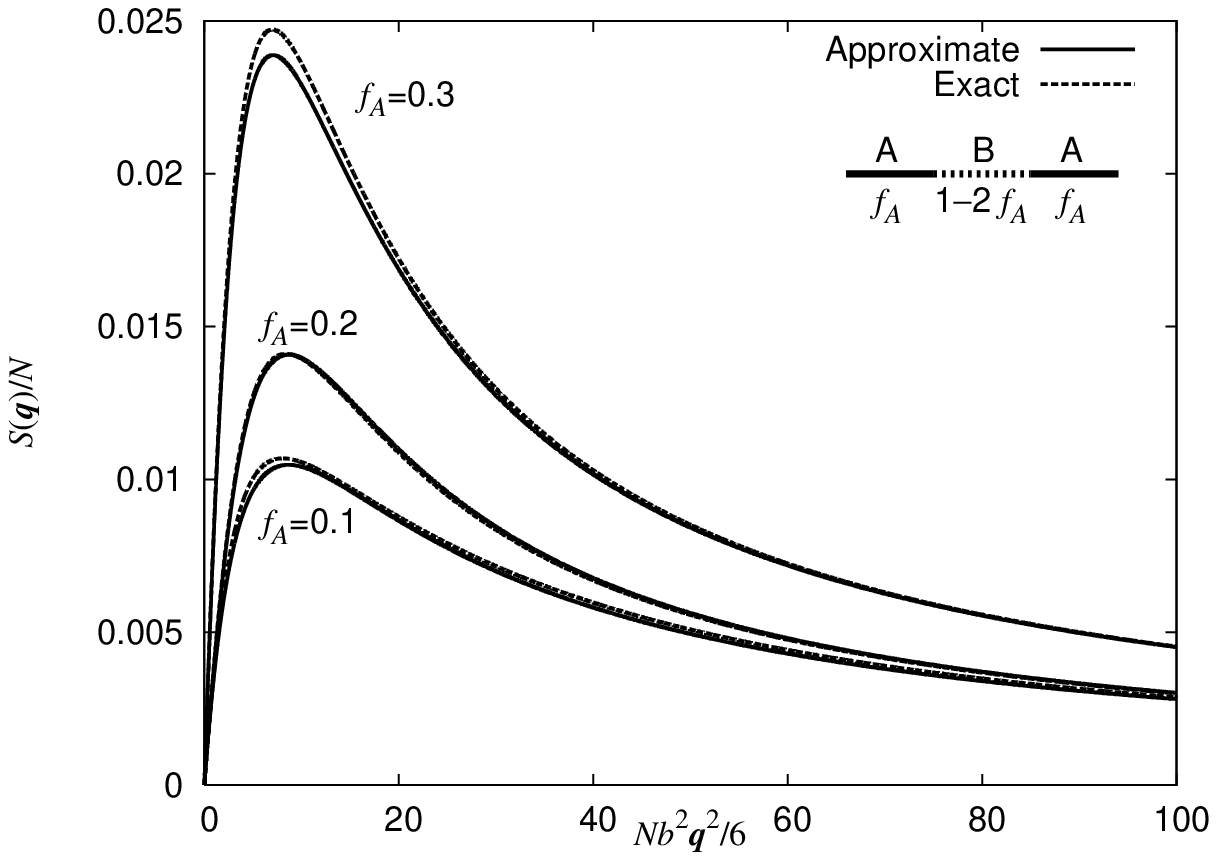}
 \caption{}
 \label{S_triblock}
\end{figure}

\begin{figure}[htbp]
 \centering
 \includegraphics{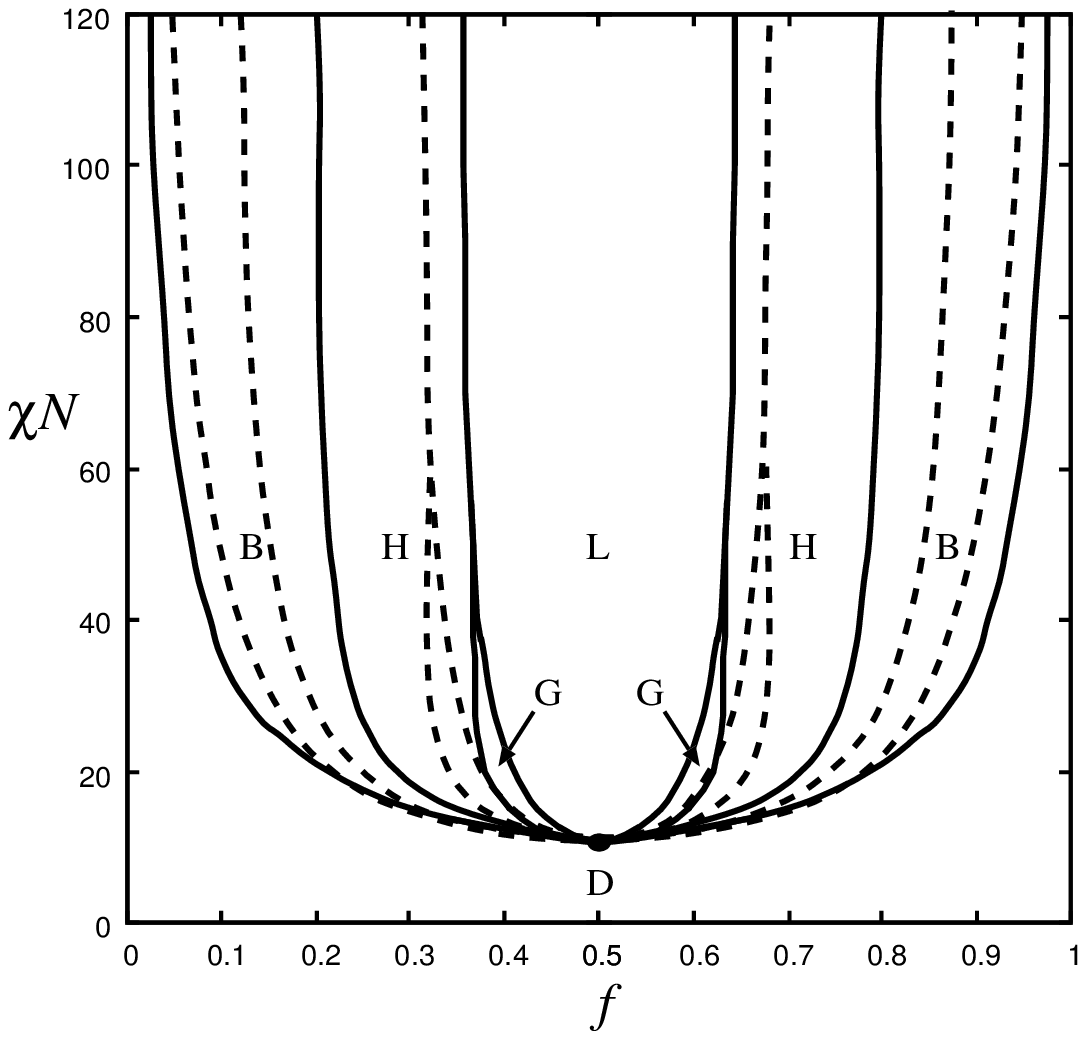}
 \caption{}
 \label{AB_phasediagram}
\end{figure}

\begin{figure}[htbp]
 \centering
 \includegraphics{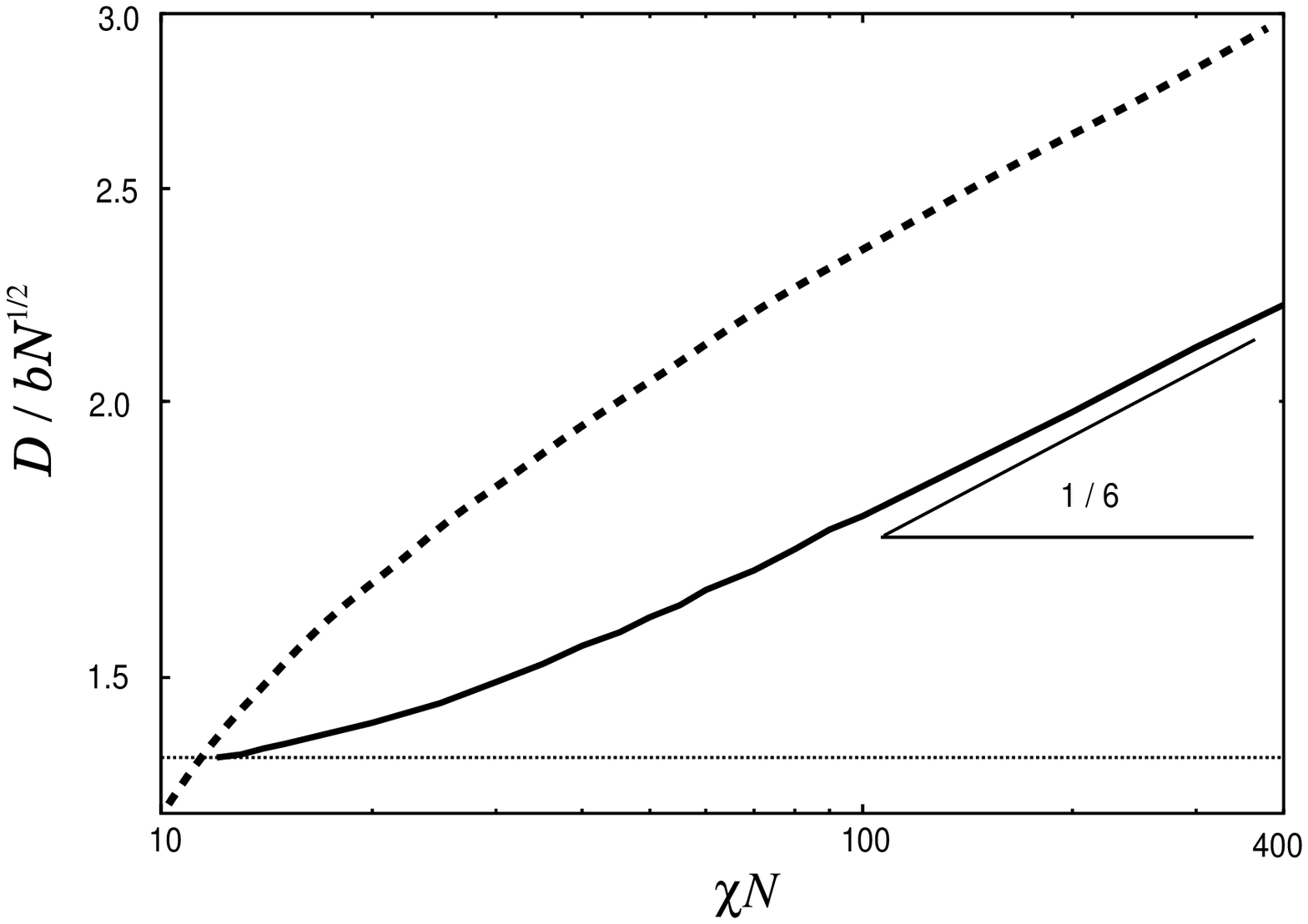}
 \caption{}
 \label{AB_period}
\end{figure}

\begin{figure}[htbp]
 \centering
 \includegraphics{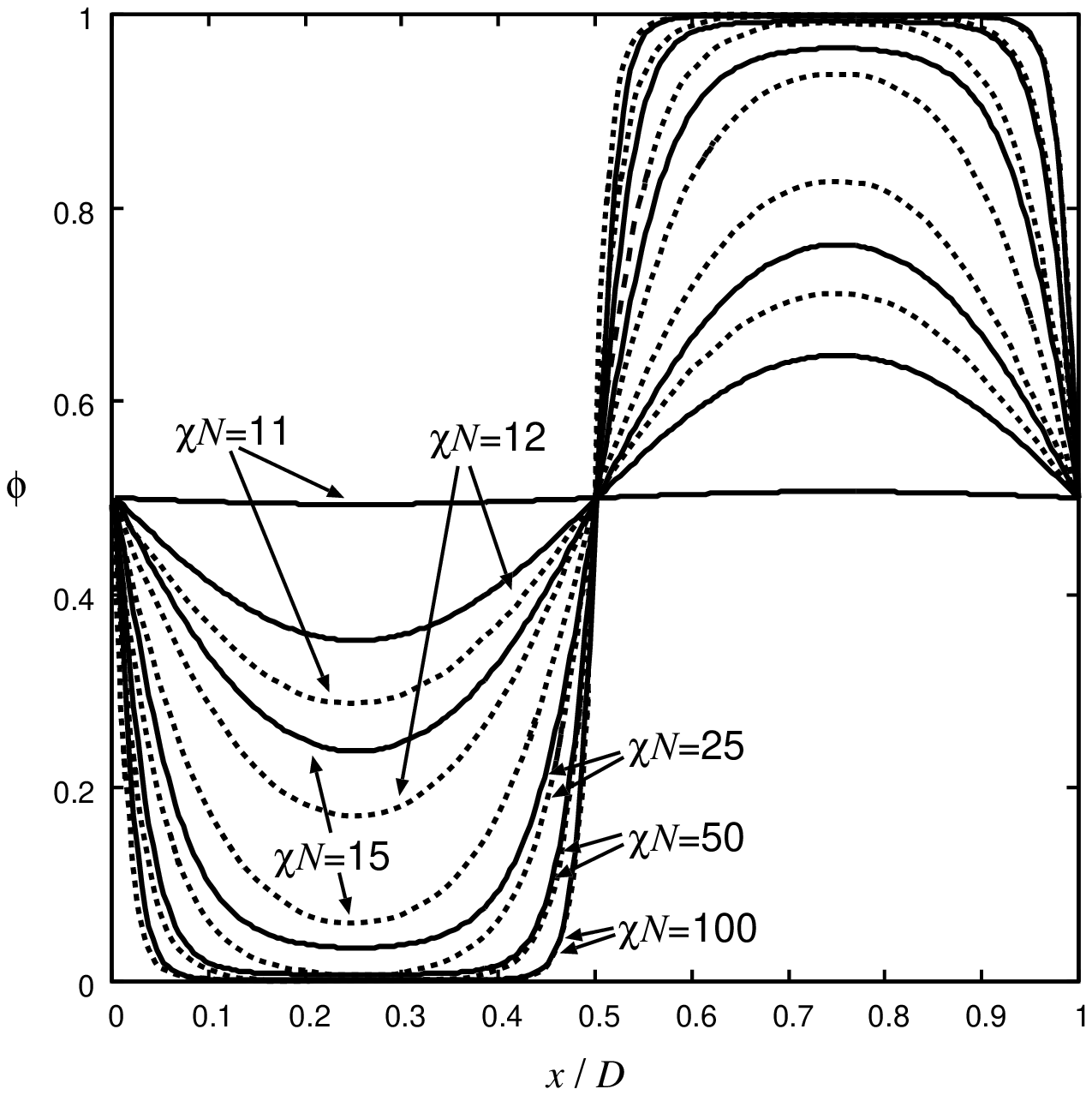}
 \caption{}
 \label{AB_equilibrium}
\end{figure}

\begin{figure}[htbp]
 \centering
 \includegraphics{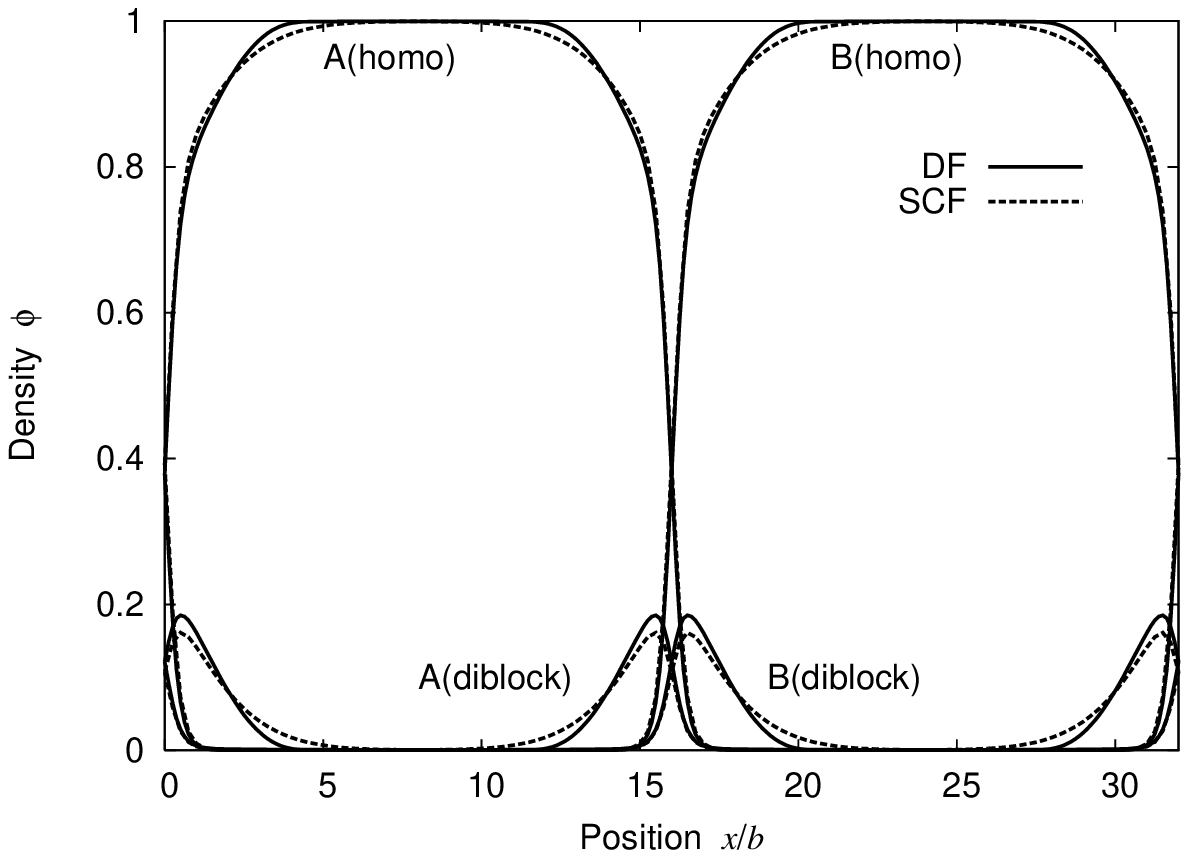}
 \caption{}
 \label{A_B_AB_equilibrium}
\end{figure}

\begin{figure}[htbp]
 \centering
 \includegraphics{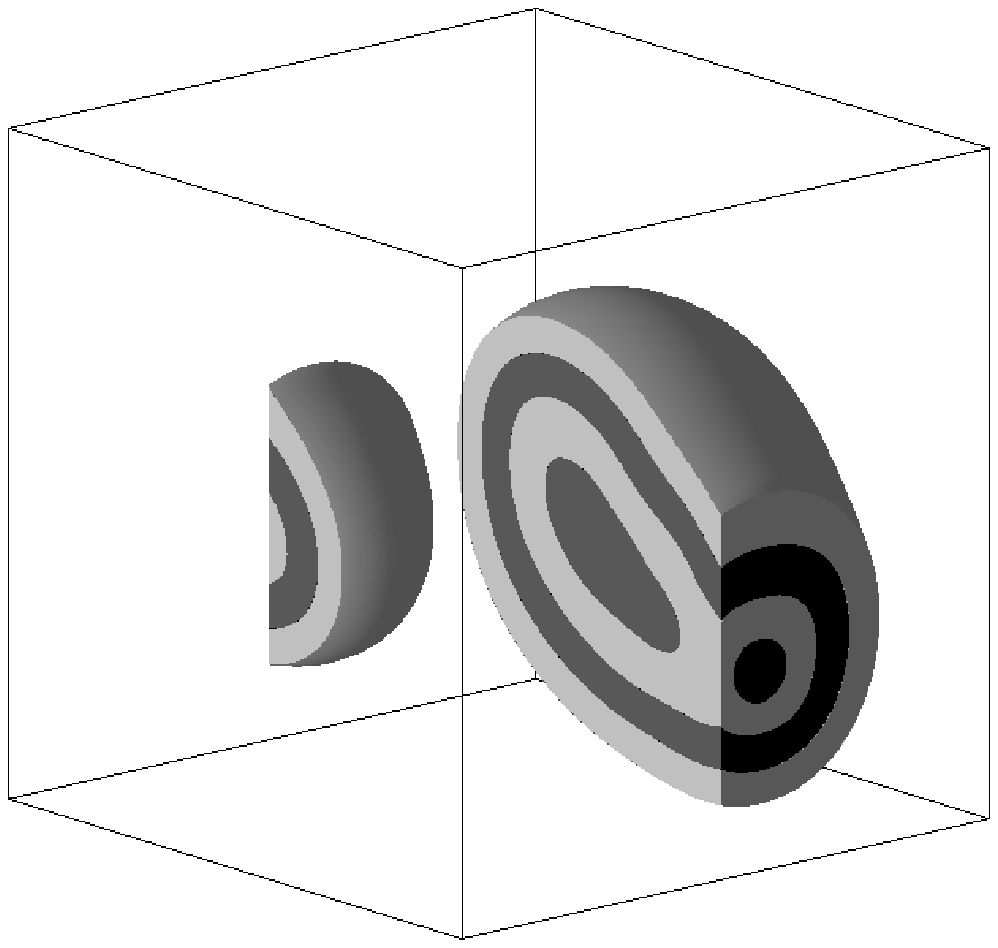}
 \caption{}
 \label{AB_C_equilibrium1}
\end{figure}

\begin{figure}[htbp]
 \centering
 \includegraphics{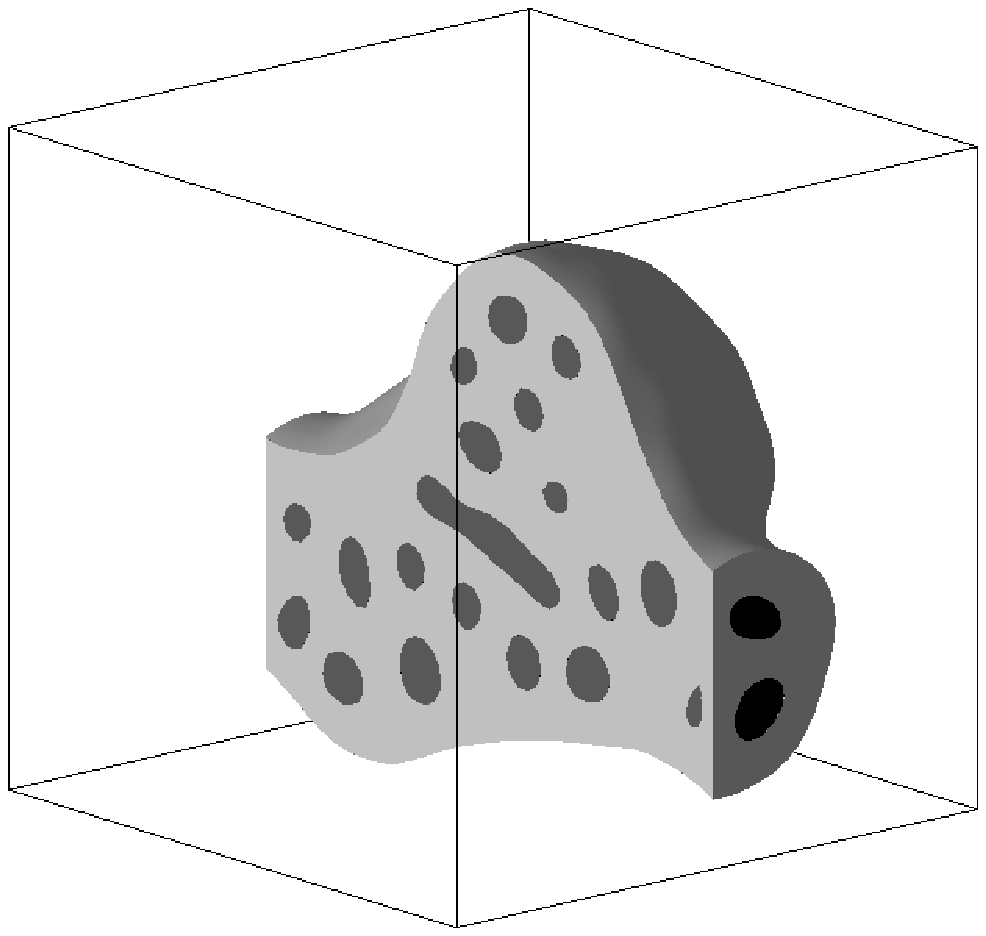}
 \caption{}
 \label{AB_C_equilibrium2}
\end{figure}

\clearpage


\newpage

\hspace{-\parindent}Figure \ref{mean_square_distance}:
 {Chemical distance $M_{p,ij}$ of block copolymers}
\vspace{\baselineskip}

\hspace{-\parindent}Figure \ref{S_diblock}:
 {Scattering functions of AB diblock copolymer melts. The $\chi$
 parameter $\chi_{AB}$ is set to zero, and $f_A$ stands for the block
 ratio.}
\vspace{\baselineskip}

\hspace{-\parindent}Figure \ref{S_triblock}:
 {Scattering functions of symmetric ABA triblock polymer melts. 
 The $\chi$ parameter $\chi_{AB}$ is set to zero, and $f_A$ stands for the block
 ratio defined in the figure.}
\vspace{\baselineskip}

\hspace{-\parindent}Figure \ref{AB_phasediagram}:
 {Phase diagram for AB diblock copolymer melts. The
 solid lines are the results of the DF simulation and the dashed lines
 are the results of the SCF simulation \cite{Matsen-Bates-1996}. (B: BCC sphere, H: Hexagonal
 cylinder, G: Double gyroid, L: Lamellar, D: Disordered)}
\vspace{\baselineskip}

\hspace{-\parindent}Figure \ref{AB_period}:
 {Equilibrium periods for AB diblock copolymer melts. Here $D$ is the
 equilibrium periods for lamellar structure. The
 solid lines are the results of the DF simulation and the dashed lines
 are the results of the SCF simulation \cite{Matsen-Bates-1996}. The doted line
 is calculated by the DF for the weak segregation limit.}
\vspace{\baselineskip}

\hspace{-\parindent}Figure \ref{AB_equilibrium}:
 {Equilibrium structures for AB diblock copolymer melts. The
 solid line is the result of the DF simulation, the doted line is the
 analytic solution for weak segregation limit by the DF and the dashed line
 is the results of the SCF simulation \cite{Matsen-Bates-1996}.}
\vspace{\baselineskip}

\hspace{-\parindent}Figure \ref{A_B_AB_equilibrium}:
 {An equilibrium structures for an A,B homopolymer / AB diblock copolymer blend}
\vspace{\baselineskip}

\hspace{-\parindent}Figure \ref{AB_C_equilibrium1}:
 {An equilibrium structures for an AB diblock coplymer / C
 homoploymer blend. The black and gray surfaces are the isodensity
 surface ($\phi_{pi}(\bm{r}) = 0.5$) for the A and B segment, respectively.}
\vspace{\baselineskip}

\hspace{-\parindent}Figure \ref{AB_C_equilibrium2}:
 {An equilibrium structures for an AB diblock copolymer / C
 homopolymer blend. The black and gray surfaces are the isodensity
 surface ($\phi_{pi}(\bm{r}) = 0.5$) for the A and B segment, respectively.}


\end{document}